\documentclass[twocolumn,showpacs,showkeys,preprintnumbers,amssymb,aps,superscriptaddress,pre]{revtex4}
\usepackage{dcolumn}
\usepackage{bm}

\usepackage{epsfig}
\usepackage{color}

\begin{document}

\preprint{APS/123-QED}

\title{Enhanced magnetoelectric effect near a field-driven zero-temperature quantum phase transition of the spin-1/2 Heisenberg-Ising ladder}
\author{Jozef Stre\v{c}ka}
\affiliation{Institute of Physics, Faculty of Science, P.~J.~\v{S}af\'arik University, Park Angelinum 9, 040\,01 Ko\v{s}ice, Slovak Republic}
\author{Lucia G\'{a}lisov\'{a}}
\email{galisova.lucia@gmail.com}
\affiliation{Institute of Manufacturing Management, Faculty of Manufacturing Technologies with the seat in Pre\v{s}ov, Technical University of~Ko\v{s}ice, Bayerova 1, 080\,01 Pre\v{s}ov, Slovak Republic}
\author{Taras Verkholyak}
\affiliation{Institute for Condensed Matter Physics, National Academy of Sciences of Ukraine, Svientsitskii Street 1, 790\,11 L'viv, Ukraine}

\date{\today}

\begin{abstract}
Magnetoelectric effect of the spin-1/2 Heisenberg-Ising ladder in a presence of the external electric and magnetic fields is rigorously examined by taking into account Katsura-Nagaosa-Balatsky mechanism. It is shown that the applied electric field may control a quantum phase transition between the N\'eel (stripy) ordered phase and the disordered paramagnetic phase. The staggered magnetization vanishes according to a power law with the Ising-type critical exponent 1/8, the electric polarization exhibits a weak singularity and the dielectric susceptibility shows a logarithmic divergence at this particular quantum phase transition. The external electric field may alternatively invoke a discontinuous phase transition accompanied with abrupt jumps of the dielectric polarization and susceptibility on assumption that the external magnetic field becomes nonzero. 
\end{abstract}

\pacs{05.50.-d, 75.10.Jm, 75.40.Cx, 75.80.+q}
\keywords{Heisenberg-Ising ladder, magnetoelectric effect, quantum phase transition, exact results}

\maketitle

\section{Introduction}

Although early investigations of the magnetoelectric effect (MEE) date back to 19th century~\cite{Fri05}, this phenomenon is subject of the renewed research interest mainly due to its wide application potential in modern technologies~\cite{Wan14}. A dependence of the magnetization on an electric field and the electric polarization on a magnetic field can be described by several alternative mechanisms. 
According to Katsura-Nagaosa-Balatsky (KNB) mechanism~\cite{Kat05} the dielectric polarization $\mathbf{p}_{i,j}$ is connected to the spin current $\mathbf{j}_{i,j}$ between a pair of the neighboring spins $\mathbf{s}_{i}$ and $\mathbf{s}_{j}$ through the expression:
\begin{equation}
\label{eq:Pij}
\mathbf{p}_{i,j} \propto  \mathbf{e}_{i,j}\times \mathbf{j}_{i,j},
\end{equation} 
where $\mathbf{e}_{i,j}$ is the unit vector pointing from $i$th to $j$th lattice site and the spin current $\mathbf{j}_{i,j}\propto \mathbf{s}_{i}\times\mathbf{s}_{j}$ is proportional to the antisymmetric Dzyaloshinskii-Moriya (DM) term \cite{Dzy58,Mor60} on assumption that two neighboring spins are coupled through the isotropic exchange interaction. 

An experimental observation of the striking magnetoelectric response of certain quasi-one-dimensional magnetic materials is another reason for a current substantial effort aimed at a more comprehensive understanding of the MEE conditioned by the KNB mechanism \cite{Yas08,Sek10}. Up to now, rigorous studies of the MEE arising from the KNB mechanism are limited to a few paradigmatic quantum spin chains such as the XXZ Heisenberg chain~\cite{Bro13}, the XY chain with three-spin interaction~\cite{Men15,Szn18}, the XY zig-zag chain~\cite{Bar18}, and the quantum compass chain~\cite{You14}. 

The main goal of the present work is to extend the aforementioned class of the one-dimensional quantum spin models. For this purpose, we will examine the spin-1/2 Heisenberg-Ising ladder simultaneously in a longitudinal magnetic field and an electric field applied along the $y$-axis in the space. As we demonstrate hereafter, the proposed quantum spin model represents a quite well tool for a rigorous study of the enhanced MEE conditioned by the KNB mechanism near the continuous quantum phase transion in the ground state.

The outline of the paper is as follows. In Sec.~\ref{sec:2} we define the quantum model and briefly metion basic steps leading to a rigorous solution of its ground state. The most interesting numerical results dealing with the MEE will be presented in Secs.~\ref{sec:3} and~\ref{sec:4}. Finally, some summarized ideas are posted in the Sec.~\ref{sec:4}. 

\section{Model and its solution}
\label{sec:2}

\begin{figure}[ht]
	\begin{center}
		\hspace{0.5cm}
		\includegraphics[width=1.0\columnwidth]{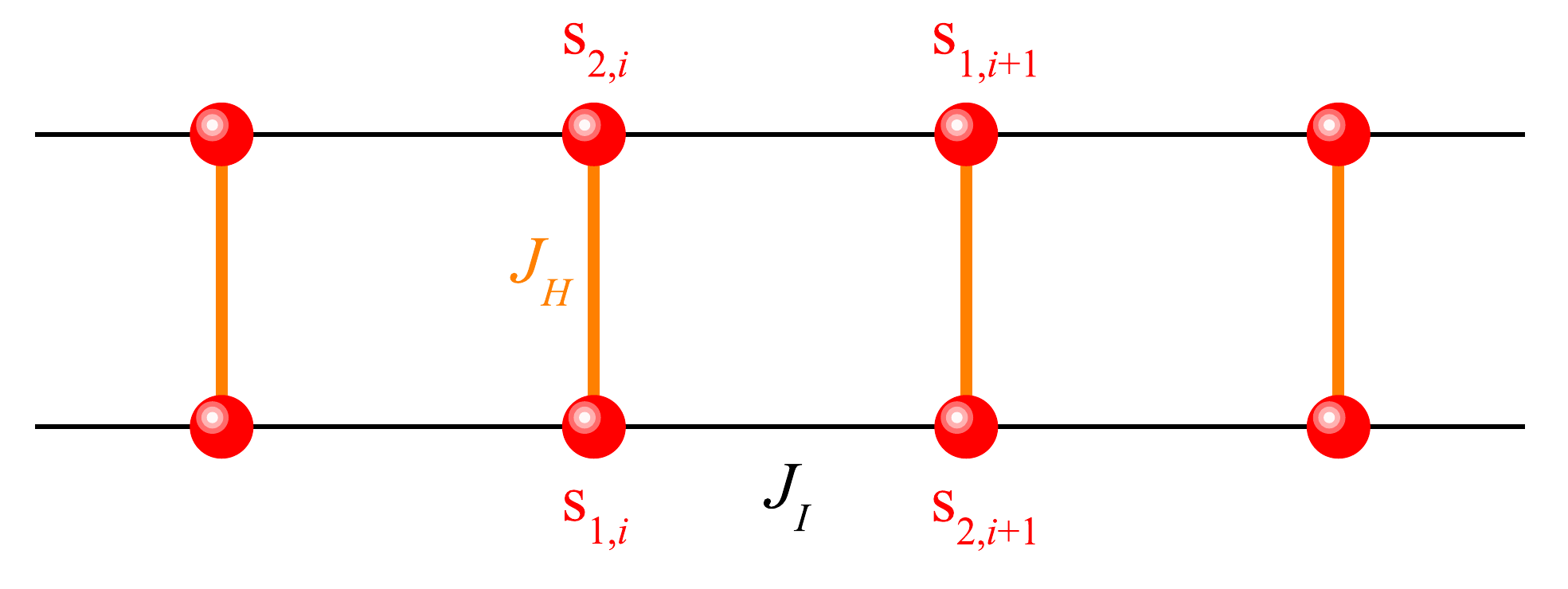}
		\vspace{-0.5cm}
		\caption{\small (Color online) A part of the quantum spin-1/2 Heisenberg-Ising two-leg ladder. Thick orange (thin black) lines represent the Heisenberg intra-rung (the Ising intra-leg) interactions in the system.}
		\label{fig1}
	\end{center}
	\vspace{0.0cm}
\end{figure}

Let us consider the quantum spin-$1/2$ Heisenberg-Ising ladder defined through the Hamitonian (see Fig.~\ref{fig1}):
\begin{eqnarray}
\label{eq:H_tot1}
{\cal H}\!\!&=&\!\!
\sum_{i = 1}^{N} 
\Big[
J_{H}\,{\mathbf s}_{1,i}\cdot{\mathbf s}_{2,i} +
J_{I}\left(s_{1,i}^{z}s_{1,i+1}^{z} + s_{2,i}^{z}s_{2,i+1}^{z}\right)
\nonumber\\
&&\qquad 
-\, h\left(s_{1,i}^{z} + s_{2,i}^{z}\right)
-E\left(s_{1,i}^{y}s_{2,i}^{x}-s_{1,i}^{x}s_{2,i}^{y}\right) 
\!
\Big],
\end{eqnarray}
where ${\mathbf s}_{l,i} \equiv (s_{l,i}^{x}, s_{l,i}^{y}, s_{l,i}^{z})$ represents the standard spin-1/2 operator for the $i$th site of the $l$th leg ($i=1,2,\ldots,N$; $l=1,2$), the parameter $J_{H}>0$ denotes the antiferromagnetic Heisenberg intra-rung interaction, 
$J_I>0$ ($J_I<0$) labels the antiferromagnetic (ferromagnetic) Ising intra-leg interaction, and $N$ denotes the total number of ladder's rungs under the periodic boundary condition ${\mathbf s}_{l,N+1}\equiv {\mathbf s}_{l,1}$. 
Finally, the last two terms in Eq.~(\ref{eq:H_tot1}) represent the standard Zeeman term associated with the magnetic field $h$ applied along the $z$-axis and the inverse DM term connected to the electric field $E$ applied along the $y$-axis, which affects the corresponding component of the dielectric polarization~(\ref{eq:Pij}), respectively. By supposing that ladder's rungs are aligned along the $x$-axis in the space, i.e. $\mathbf{e}_{(1,i),(2,i)} = (1,0,0)$ in Eq.~(\ref{eq:Pij}), the dielectric polarization $p_i^y=s_{1,i}^{y}s_{2,i}^{x}-s_{1,i}^{x}s_{2,i}^{y}$ is prescribed to the $i$th rung. Note that the dielectric dipole moment is imposed to be generated on the Heisenberg rungs only, whereas the Ising bonds along legs does not exhibit any magnetoelectric connection.

The DM term associated with the electric field can be eliminated from the Hamiltonian~(\ref{eq:H_tot1}) by performing a local spin-rotation transformation around the $z$-axis by the specific angle $\varphi = \tan^{-1}(E/J_H)$~\cite{Aff99}. As a result, one gets the Hamiltonian of the spin-1/2 Heisenberg-Ising ladder with the effective XXZ intra-rung interaction and the Ising intra-leg interaction in a magnetic field:
\begin{eqnarray}
\label{eq:H_tot2}
{\cal H}\!\!&=&\!\!
\sum_{i = 1}^{N} 
\Big[
\sqrt{J_H^2+E^2}\left(s_{1,i}^{x}s_{2,i}^{x}+s_{1,i}^{y}s_{2,i}^{y}\right)+J_{H}s_{1,i}^{z}s_{2,i}^{z} \nonumber\\
&&
+\,
J_{I}\left(s_{1,i}^{z}s_{1,i+1}^{z} + s_{2,i}^{z}s_{2,i+1}^{z}\right)
- h\left(s_{1,i}^{z} + s_{2,i}^{z}\right)
\!
\Big],
\end{eqnarray}
which can be alternatively viewed as a special case of the frustrated spin-1/2 Heisenberg-Ising ladders exactly solved in the recent works~\cite{Ver12,Ver13,Brze08}. Therefore, it is sufficient to closely follow the approach elaborated in Ref.~\cite{Ver12} to get a rigorous solution for the ground state of the investigated quantum spin model. 
First we define the eigenstates of the XXZ bonds on the rungs:
\begin{eqnarray}
\label{eq:bond_st}
&& |\phi_{0,\pm}^{i}\rangle=\frac{1}{\sqrt2}
(|\! \downarrow_{1,i}\uparrow_{2,i}\rangle \pm |\! \uparrow_{1,i}\downarrow_{2,i}\rangle),
\nonumber\\
&& |\phi_{1,\pm}^{i}\rangle=\frac{1}{\sqrt2}
(|\! \uparrow_{1,i}\uparrow_{2,i}\rangle \pm |\! \downarrow_{1,i}\downarrow_{2,i}\rangle),
\end{eqnarray}
where the states $|\phi_{0,\pm}^{i}\rangle$ ($|\phi_{1,\pm}^{i}\rangle$) belong to the subspace with $(S^z_i)^2=(s_{1,i}^{z}+s_{2,i}^{z})^2=0$ (1). In the next step we introduce the pseudo-spin notations for the bond states as 
$|\phi_{0,-}^{i}\rangle=|\! \! \downarrow\rangle_0^i$, $|\phi_{0,+}^{i}\rangle=|\! \! \uparrow\rangle_0^i$ (for $(S_i^z)^2=0$ subspace), 
and $|\phi_{1,-}^{i}\rangle=|\! \! \downarrow\rangle_1^i$, $|\phi_{1,+}^{i}\rangle=|\! \! \uparrow\rangle_1^i$ (for $(S_i^z)^2=1$ subspace). 
We also need to set a new spin operators $\tilde{s}_{i}^\alpha\,(\alpha = x,\,y,\,z)$ acting on the pseudo-spin basis as well as the binary variable $n_i = 0$ and $1$ assigned to $(S_i^z)^2=0$ and $1$ states of the $i$th rung (dimer), respectively. By straightforward calculation one can establish the following relations for the spin operators:
\begin{eqnarray}
\label{eq:pseudo-spins}
s_{1,i}^{z} \!\!&=&\!\! (2n_{i}-1)\tilde{s}_{i}^{x},\quad s_{2,i}^{z} = \tilde{s}_{i}^{x}, 
\nonumber\\
s_{1,i}^{x}s_{2,i}^{x} \!\!&=&\!\!  \frac{\tilde{s}_{i}^{z}}{2}, 
\,
s_{1,i}^{y}s_{2,i}^{y} =  (1-2n_{i})\frac{\tilde{s}_{i}^{z}}{2}, 
\,
s_{1,i}^{z}s_{2,i}^{z} = \frac{1}{4}(2n_{i}-1).
\nonumber
\end{eqnarray}
Consequently, it results in the following equivalent pseudo-spin representation of the Hamiltonian~(\ref{eq:H_tot2}): 
\begin{eqnarray}
\label{eq:H_tot3}
{\cal H} \!\!&=&\!\! \sum_{i = 1}^{N} \! \Big\{
2J_I\left[n_{i}n_{i+1} + (1-n_{i})(1-n_{i+1})\right]\tilde{s}_{i}^{x}\tilde{s}_{i+1}^{x}
\nonumber\\
\!\!&+&\!\! \sqrt{J_{H}^2{+}E^2}(1-n_i)\tilde{s}_{i}^{z} {-} 2hn_i\tilde{s}_{i}^{x} {+} \frac{J_H}{4}(2n_{i}{-}1)  \Big\}. 
\end{eqnarray}

It has been verified in Ref.~\cite{Ver12} that the lowest-energy eigenstates derived from the effective Hamiltonian~(\ref{eq:H_tot3}) of the spin-1/2 Heisenberg-Ising ladder in zero magnetic field ($h = 0$) follow from two exactly solved spin-chain models, namely, the spin-1/2 Ising chain in a transverse magnetic field~\cite{Pfe70} acquired from the effective Hamiltonian~(\ref{eq:H_tot3}) by assuming all rung states are in the $(S_i^z)^2=0$ subspace ($n_{i}=0$) and the spin-1/2 Ising chain in a longitudinal magnetic field~\cite{Bax82} acquired from the effective Hamiltonian~(\ref{eq:H_tot3}) by considering all rung states in the $(S_i^z)^2=1$ subspace ($n_{i}=1$). In a nonzero magnetic field ($h \neq 0$) one additionally has to consider another exactly solvable effective spin-chain model acquired from the effective Hamiltonian~(\ref{eq:H_tot3}) by assuming a regular alternation of the singlet and triplet states on odd and even rungs that correspond to $n_{2i-1} = 0$ and $n_{2i} = 1$ or vice versa \cite{Ver12}. From this point of view, the exact solution for a ground state of the spin-1/2 Heisenberg-Ising ladder in electric and magnetic fields defined through the Hamiltonian~(\ref{eq:H_tot1}) is  formally completed.

\section{MEE in zero magnetic field}
\label{sec:3}

\begin{figure*}[ht!]%
	\centering
	\includegraphics*[width=1.0\textwidth]{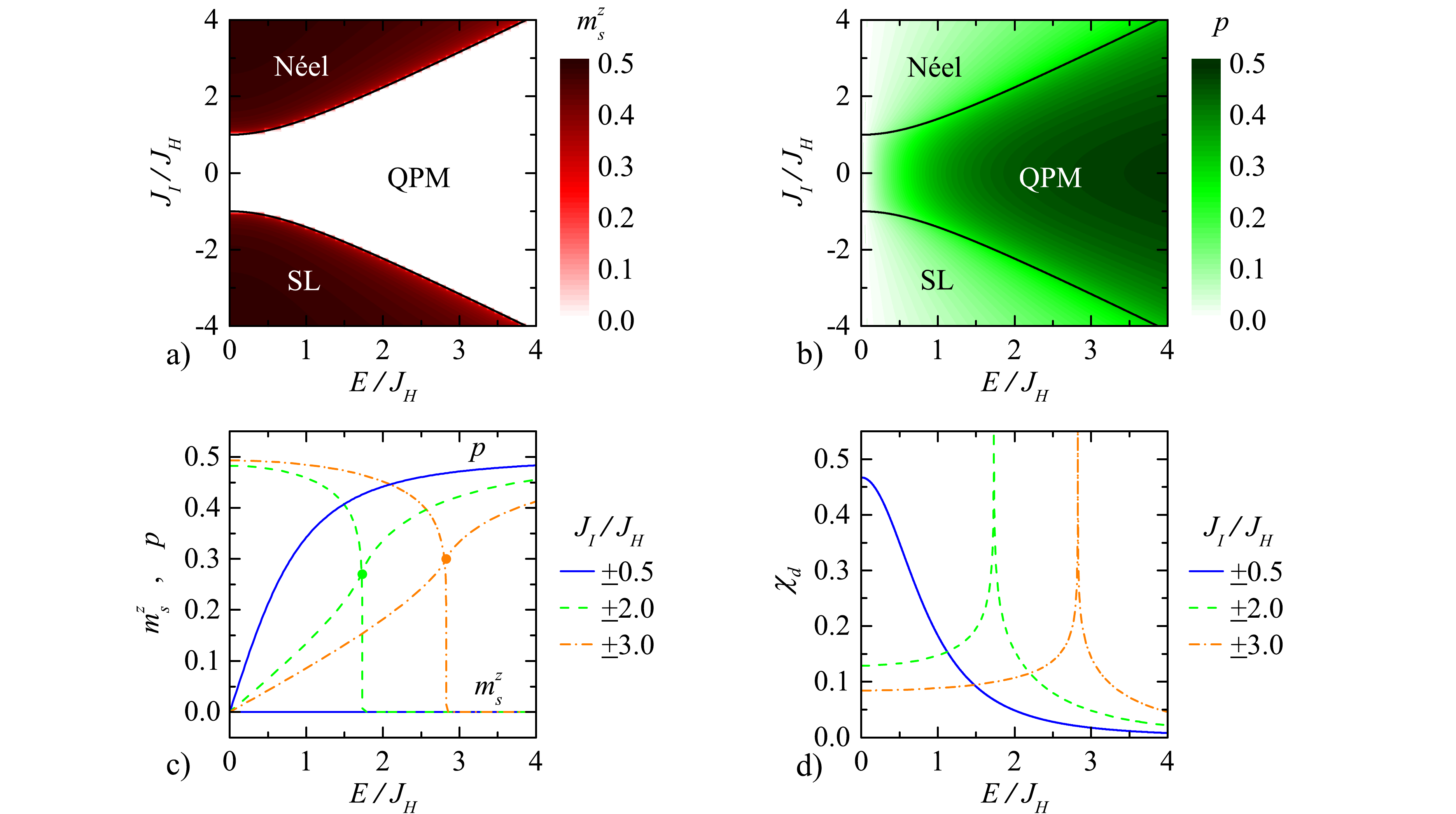}
	\vspace{-0.35cm}
	\caption{(Color online) The density plot of the staggered magnetization $m_s^z$ (panel a) and the dielectric polarization $p$ (panel~b) as ground-state phase diagrams in the $E/J_H-J_I/J_H$ plane together with the electric-field variations of the staggered magnetization $m_s^z$, the dielectric polarization $p$ (panel c) and the dielectric susceptibility $\chi_d$ (panel d) for the interaction ratios $J_I/J_H = \pm0.5$, $\pm2.0$, $\pm3.0$.}
	\label{fig2}
\end{figure*}

The ground-state energy of the spin-1/2 Heisenberg-Ising ladder in zero magnetic field ($h=0$) follows from the formula:
\begin{eqnarray}
\label{eq:e_GS}
{\cal E}_{GS} \equiv \frac{1}{N} \langle {\cal H} \rangle =  
- \frac{J_H}{4} -\frac{\sqrt{J_{H}^2+E^2}+|J_I|}{\pi}\mathbf{E}(a), 
\end{eqnarray}
where $\mathbf{E}(a) \!\!\!=\!\!\!\! \int_{0}^{\!\frac{\pi}{2}}\!{\rm d}\phi\sqrt{1\!-\!a^2\sin^2\!\phi}$ is the complete elliptic integral of the second kind with $a^2= \frac{4 |\lambda|}{(1 + |\lambda|)^2}$ and $\lambda = \frac{J_{I}}{\sqrt{J_{H}^{2}+E^{2}}}$. The ground-state energy (\ref{eq:e_GS}) has a singularity at $|\lambda|=1$ ($|E_c|=\sqrt{J_I^2-J_H^2}$\,), which relates to a quantum phase transition between the quantum paramagnetic (QPM) phase emergent for $|\lambda| \leq 1$ and either N\'eel or stripe-leg (SL) phase emergent for $|\lambda|>1$ depending on whether $J_I>0$ or $J_I<0$, respectively. While the N\'eel and stripy spin orders are quite analogous and can be characterized through the nonzero staggered magnetization, this order parameter becomes zero within the disordered QPM phase:  
\begin{eqnarray}
\label{ms}
m_{s}^{z} \equiv \frac{1}{2} \langle |s_{1,i}^z-s_{2,i}^z| \rangle = \left\{ 
\begin{array}{ll}
\displaystyle\frac{1}{2}\!\left(1 - \frac{1}{\lambda^{2}}\right)^{\!\!1/8} & \textrm{ if } |\lambda|>1
\\[3mm]
\displaystyle 0 & \textrm{ if } |\lambda|\leq 1
\end{array} 
\right.\!\!.
\end{eqnarray}
Obviously, the staggered magnetization $m_s^z$ displays a steep power-law decline with the Ising-type critical exponent $\beta=1/8$ at the quantum phase transition $|\lambda|=1$ (see Figs.~\ref{fig2}a,\,c). Contrary to this, the dielectric polarization is governed by the formula:  
\begin{eqnarray}
\label{p}
p \equiv \langle p_i^y \rangle = \frac{1}{2\pi}\frac{E}{\sqrt{J_{H}^{2}{+}E^{2}}}
[(1{+}|\lambda|)\mathbf{E}(a) {+} (1{-}|\lambda|)\mathbf{K}(a)],
\end{eqnarray}
where $\mathbf{K}(a) {=} \int_{0}^{\!\frac{\pi}{2}}\!{\rm d}\phi(1{-}a^2\sin^2\!\phi)^{-1/2}$ is the complete elliptic integral of the first kind. The formula~(\ref{p}) implies a smoother change of the dielectric polarization with only a weak-type singularity $|p - p_c|\sim (E-E_c) \ln|E-E_c|$ when crossing the respective ground-state phase boundary (see Figs.~\ref{fig2}b,\,c). Note furthermore that the dielectric polarization is much higher in the disordered QPM phase than in the ordered N\'eel and SL phases as evidenced by a density plot displayed in Fig.~\ref{fig2}b. The weak singularity of $p$ at the quantum critical point (a filled circle in Fig.~\ref{fig2}c) is more markedly evidenced through a logarithmic divergence of the dielectric susceptibility $\chi_{d} = \frac{\partial p}{\partial E}\sim -\ln|E-E_c|$ at the critical electric field $E_c$, as shown in Fig.~\ref{fig2}d.

\begin{figure*}[ht!]%
	\centering
	\includegraphics*[width=1.0\textwidth]{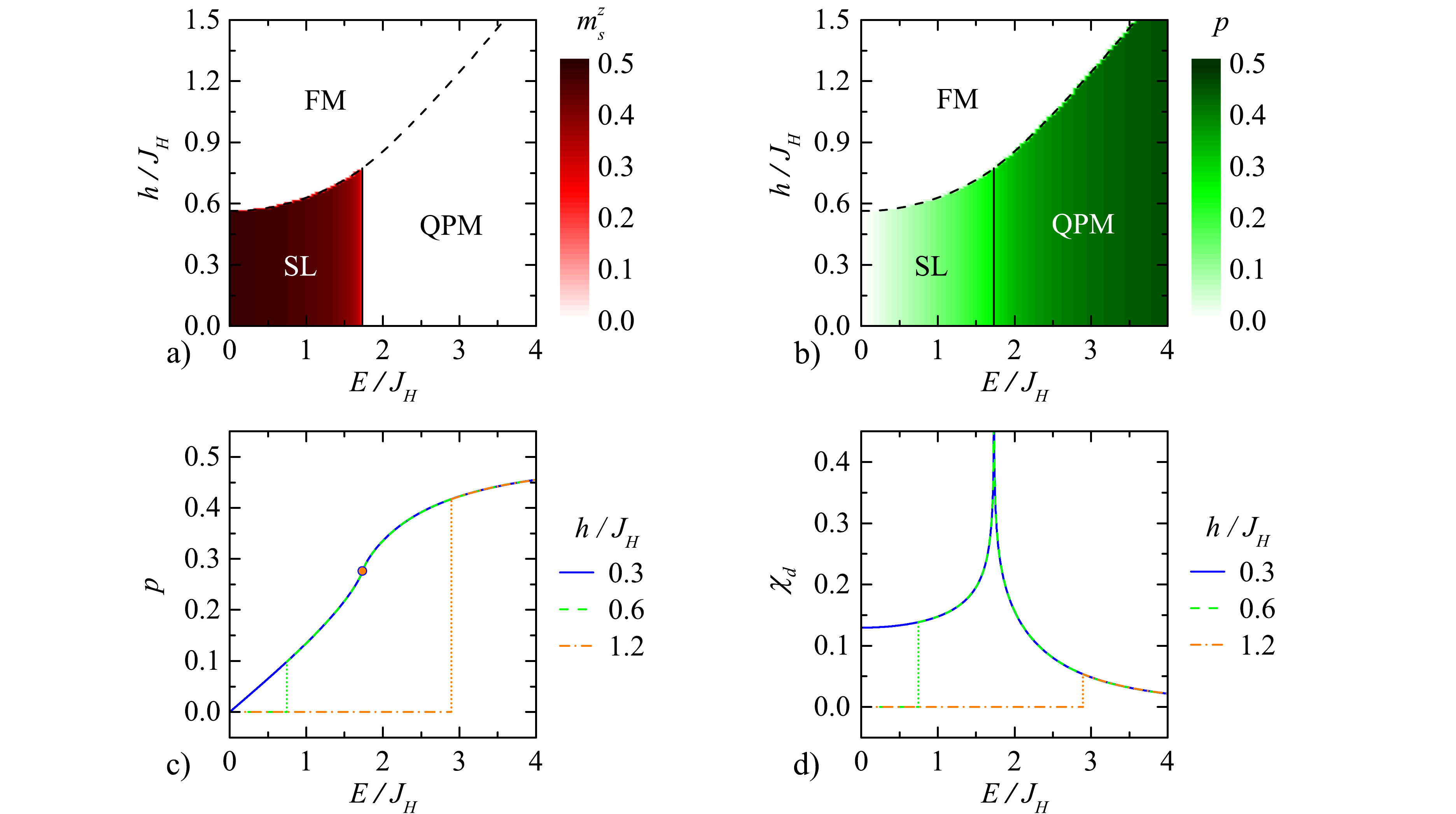}
	\vspace{-0.35cm}
	\caption{(Color online) The density plots of the staggered magnetization $m_s^z$ (panel a) and the dielectric polarization $p$ (panel~b) as ground-state phase diagrams in the $E/J_H-h/J_H$ plane for the interaction ratio $J_I/J_H = -2$ along with the electric-field variations of the dielectric polarization $p$ (panel c) and the susceptibility $\chi_{d}$ (panel c) for the interaction ratio $J_I/J_H = -2$ and three different magnetic fields $h/J_H = 0.3$, $0.6$, $1.2$.}
	\label{fig3}
\end{figure*}

\section{MEE in nonzero magnetic field}
\label{sec:4}

Another two ground states may emerge due to nonzero magnetic field ($h\neq0$) . At high enough magnetic fields one may detect the classical ferromagnetic (FM) phase characterized by: 
\begin{eqnarray}
{\cal E}_{\rm FM} = \frac{J_{H}}{4} + \frac{J_{I}}{2} - h; \quad m_{s}^{z} = 0; \quad p = 0; \quad \chi_d = 0.
\end{eqnarray}
In addition, the staggered bond (SB) phase with a regular alternation of the singlet and polarized triplet states characterized by:
\begin{eqnarray}
{\cal E}_{\rm SB} &=& -\frac{1}{4}\sqrt{J_{H}^{2}+E^{2}} - \frac{h}{2};  \quad  m_{s}^{z} = \frac{1}{4}; \nonumber\\
p &=& \frac{1}{4}\frac{E}{\sqrt{J_{H}^{2}+E^{2}}}; \quad \chi_d = \frac{1}{4}\frac{J_{H}^{2}}{\left(J_{H}^{2}+E^{2}\right)^{3/2}}
\end{eqnarray}
may appear at moderate magnetic fields provided that the Ising intra-leg coupling is of the antiferromagnetic nature $J_I > 0$. The typical ground-state phase diagram of the spin-1/2 Heisenberg-Ising ladder with the ferromagnetic Ising intra-leg coupling is illustrated in Fig.~\ref{fig3} in a form of density plots of the staggered magnetization $m_s^z$ (Fig.~\ref{fig3}a) and the dielectric polarization $p$ (Fig. \ref{fig3}b) in the $E/J_H-h/J_H$ plane. Obviously, Figs.~\ref{fig3}a,\,b repeatedly imply a competitive character of the magnetic and dielectric spin orders when an enhancement of the dielectric polarization is accompanied with a reduction of the staggered magnetization or vice versa. 

Furthermore, Figs. \ref{fig3}c,\,d display a few typical variations of the dielectric polarization and susceptibility across continuous and discontinuous phase transitions, which may be apparently controlled by the external electric and magnetic fields. For sufficiently low magnetic fields (e.g. $h/J_H = 0.3$) the dielectric polarization displays a continuous rise with a weak singularity at the quantum phase transition between the SL and QPM phases (a filled circle in Fig.~\ref{fig3}c), at which the dielectric susceptibility diverges logarithmically, as shown in Fig.~\ref{fig3}d. Contrary to this, the dielectric polarization and susceptibility may exhibit abrupt jumps related to a discontinuous phase transition between the FM and SL phases at moderate magnetic fields (e.g. $h/J_H = 0.6$) before achieving the electric-field-driven quantum phase transition SL--QPM. 

\begin{figure*}[htb]%
	\centering
	\includegraphics*[width=1.0\textwidth]{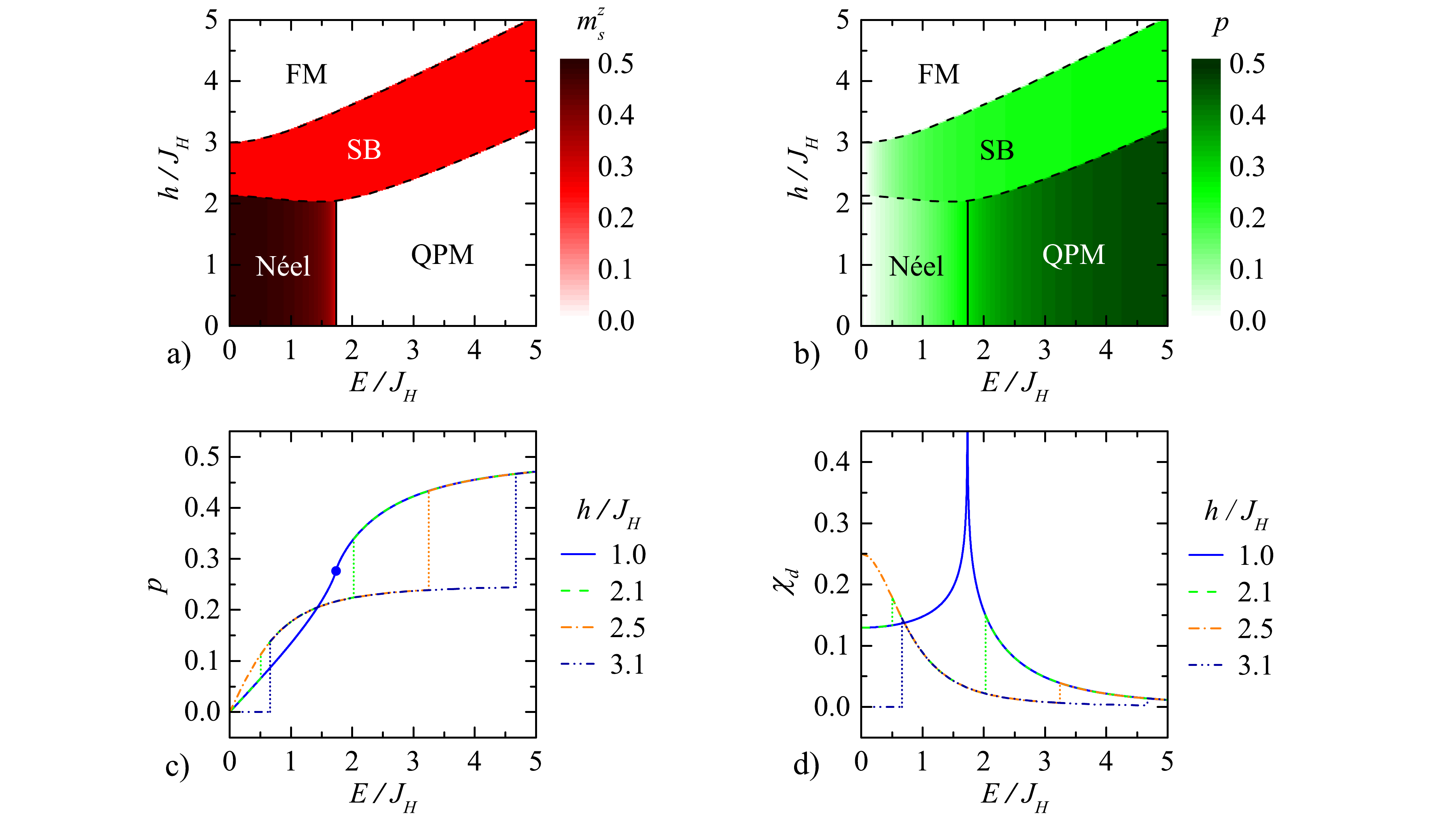}
	\vspace{-0.35cm}
	\caption{(Color online) The density plots of the staggered magnetization $m_s^z$ (panel~a) and the dielectric polarization $p$ (panel~b) as ground-state phase diagrams in the $E/J_H-h/J_H$ plane for the interaction ratio $J_I/J_H = 2$ along with the electric-field variations of the dielectric polarization $p$ (panel c) and the susceptibility $\chi_d$ (panel d) for the interaction ratio $J_I/J_H = 2$ and four different magnetic fields $h/J_H = 1.0$, $2.1$, $2.5$, $3.1$.}
	\label{fig4}
\end{figure*}

To compare with, the typical ground-state phase diagram of the spin-1/2 Heisenberg-Ising ladder with the antiferromagnetic Ising intra-leg coupling is displayed in Fig.~\ref{fig4} in the form of the density plots of the staggered magnetization $m_s^z$ (Fig.~\ref{fig4}a) and the dielectric polarization $p$ (Fig.~\ref{fig4}b) in the $E/J_H-h/J_H$ plane. Evidently, the main qualitative difference with respect to the previous case lies in a presence of the SB phase at moderate magnetic fields. Owing to this fact, one may detect much greater versatility of the electric-field variations of the dielectric polarization and susceptibility, as demonstrated in Figs.~\ref{fig4}c,\,d. At low enough magnetic fields (e.g. $h/J_H = 1.0$), the dielectric polarization exhibits a smooth continuous rise upon strengthening of the electric field with a weak singularity at the respective quantum critical point (a filled circle in Fig.~\ref{fig4}c), which becomes more evident through a logarithmic divergence of the dielectric susceptibility (see Fig.~\ref{fig4}d). By contrast, one may detect a remarkable dependence of the dielectric polarization on the electric field with either one or two discontinuous jumps, which relate to discontinuous phase transitions driven by the external electric field at moderate and high magnetic fields (e.g. $h/J_H = 2.1$, $2.5$ and $3.1$).

\section{Conclusions}
\label{sec:5}

To conclude, in the present paper, we have exactly examined a ground state of the spin-1/2 Heisenberg-Ising ladder in the external electric and magnetic fields. It has been demonstrated that the ground-state spin arrangements may be basically manipulated through the MEE conditioned by KNB mechanism via the external electric field, which additionally affords an alternative tool to control a quantum phase transition between the N\'eel (or stripy) quantum ordered phase and disordered quantum paramagnetic phase at zero magnetic field. It turns out, moreover, that an interplay between the electric and magnetic fields may cause an existence of either one or two discontinuous phase transitions as well as a single continuous quantum phase transition. Although the investigated quantum spin chain does not exhibit a spontaneous multiferroic behavior, a feedback control of the magnetic spin orderings through the external electric field might be of immense technological relevance because of a wide application potential of multifunctional materials.

\begin{acknowledgments}
This work was supported by Ministry of Education, Science, Research and Sport of the Slovak Republic under the grant VEGA 1/0043/16 and by Slovak Research and Development Agency under the contract No.~APVV-16-0186.
\end{acknowledgments}


\begin{thebibliography}{20}
\bibitem{Fri05}
M.~Fiebig, J. Phys. D \textbf{38}, R123 (2005) and references therein.
\bibitem{Wan14}%
Y.~Wang, J.~Li, D.~Viehland, Mater. Today \textbf{17}, 269 (2014).
\bibitem{Kat05}%
H.~Katsura, N.~Nagaosa, and A.V.~Balatsky, Phys. Rev. Lett. \textbf{95}, 057205 (2005).
\bibitem{Dzy58}%
I. Dzyaloshinskii, J. Phys. Chem. Solids. \textbf{4}, 241 (1958).
\bibitem{Mor60}%
T. Moriya, Phys. Rev. \textbf{120}, 91 (1960).
\bibitem{Yas08}%
Y.~Yasui, Y.~Naito, K.~Sato, T.~Moyoshi, M.~Sato, and K.~Kakurai, J. Phys. Soc. Jpn. \textbf{77}, 023712 (2008).
\bibitem{Sek10}%
S.~Seki, T.~Kurumaji, S.~Ishiwata, H.~Matsui, H.~Murakawa, Y.~Tokunaga, Y.~Kaneko, T.~Hasegawa, and Y.~Tokura, Phys. Rev. B \textbf{82}, 064424 (2010).
\bibitem{Bro13}
M.~Brockmann, A.~Kl\"umper, and V.~Ohanyan, Phys. Rev. B \textbf{87}, 054407 (2013).
\bibitem{Men15}
O.~Menchyshyn, V.~Ohanyan, T.~Verkholyak, T.~Krokhmalskii, and O.~Derzhko, Phys. Rev. B \textbf{92}, 184427 (2015).
\bibitem{Szn18}
J.~Sznajd, Phys. Rev. B \textbf{97}, 214410 (2018).
\bibitem{Bar18}
O.~Baran, V.~Ohanyan, and T.~Verkholyak, Phys. Rev. B \textbf{98}, 064415 (2018).
\bibitem{You14}
W.-L.~You, G.-H.~Liu, P.~Horsch, and A.M.~Ole\'s, Phys. Rev. B \textbf{90}, 094413 (2014).
\bibitem{Aff99}
I.~Affleck and M.~Oshikawa, Phys. Rev. B \textbf{60}, 1038 (1999).
\bibitem{Ver12}
T.~Verkholyak and J.~Stre\v{c}ka, J. Phys. A: Math. Teor. \textbf{45}, 305001 (2012).
\bibitem{Ver13}
T.~Verkholyak and J.~Stre\v{c}ka, Condens. Matter Phys. \textbf{16}, 13601 (2013).
\bibitem{Brze08}
W. Brzezicki, A.\,M. Ole\'{s}, Eur. Phys. J. B \textbf{66}, 361 (2008).
\bibitem{Pfe70}
P.~Pfeuty, Ann. Phys., NY \textbf{57}, 79 (1970).
\bibitem{Bax82}
R.J. Baxter, Exactly Solved Models in Statistical Mechanics (Academic Press, New York, 1982), p.\,32--38.
\end{thebibliography}
\end{document}